\documentstyle[12pt,aasms4]{article}	
\tighten

\begin{document}

\title {Keck Imaging of Binary L Dwarfs}

\author {D.W. Koerner\altaffilmark{1}, J.Davy 
Kirkpatrick\altaffilmark{2}, M.W. McElwain\altaffilmark{1},
and N.R. Bonaventura\altaffilmark{1}}
\altaffiltext{1} 
{University of Pennsylvania, David Rittenhouse Laboratory, 
209 S. 33rd St., Philadelphia, PA 19104-6396}
\altaffiltext{2}
{Infrared Processing and Analysis Center, MS 100-22, California Institute
of Technology, Pasadena, CA 91125}
\bigskip
\bigskip

\begin{abstract}

We present Keck near-infrared imaging of three binary L 
dwarf systems, all of which are likely to be sub-stellar. 
Two are lithium dwarfs, and a third exhibits an L7 spectral type, 
making it the coolest binary known to date. All have component
flux ratios near 1 and projected physical separations between 
5 and 10 AU, assuming distances of 18 to 26 pc from recent measurements of
trigonometric parallax. These surprisingly similar binaries 
represent the sole detections of companions in ten
L dwarf systems which were analyzed in the preliminary phase
of a much larger dual-epoch imaging survey. The detection rate
prompts us to speculate that binary companions to L dwarfs are 
common, that similar-mass systems predominate, and that their 
distribution peaks at radial distances in accord both with M dwarf 
binaries and with the radial location of Jovian planets in our own 
solar system.  To fully establish these conjectures against doubts
raised by biases 
inherent in this small preliminary survey, however, will require 
quantitative analysis of a larger volume-limited sample which has 
been observed with high resolution and dynamic range.

\end{abstract}

Subject headings: stars: circumstellar matter ---
binaries: spectroscopic -- planetary systems

\section {Introduction}

L dwarfs make up a new spectral class of objects that 
are believed to have
masses near or below the hydrogen-burning limit 
(Kirkpatrick et al.\ 1999a; 1999b). Many satisfy currently
accepted criteria for identification as {\it bona fide}  
brown dwarfs (See Tinney 1999 for a review). Their 
local field detection rate in infrared sky surveys suggests they comprise a 
sizeable population which is well represented by an extension of the 
field-star mass function, $\Psi(M) \propto M^{-\alpha}$, with 
$ 1 < \alpha < 2$ (Reid et al.\ 1999). The occurrence frequency of multiplicity
among these systems is completely unknown; it is an open question as to 
whether the distribution of their 
companions matches that of M dwarfs or bears the 
stamp of a different, sub-stellar formation mechanism.

Stellar companions are detected in approximately 35\% of M dwarf systems with 
a distribution peaking at a radius in the range $3-30$ AU 
(Fischer \& Marcy 1992; Henry \& McCarthy 1993; 
Reid \& Gizis 1997). Efforts to uncover the mass and radial distribution of 
extra-solar planets around M stars are just beginning to meet with success 
and have revealed super Jovian-mass planets within a few AU of their central 
stars, consistent with results for earlier spectral types (Marcy et al.\
1998). The relationship of this population to that of binary companions 
and planetary systems like our own is a topic of current debate (Black 
1997). The true answer will not be readily apparent until a more complete 
range of mass and orbital distances has been surveyed. Ground-based imaging of
L dwarfs provides a unique piece to this puzzle, since the reduced
glare of low-luminosity primaries affords increased sensitivity to very
faint companions.

To date, very few multiple L dwarf systems have been identified. Several
L dwarf secondaries have been discovered around nearby stars 
(Becklin \& Zuckerman 1988; Rebolo et al.\ 1998; 
Kirkpatrick et al.\ 1999b).  Among a handful of binary systems believed
to be composed of two 
brown-dwarf components (e.g., Basri \& Mart\'{\i}n 1997), 
only two have primary spectral types as late as L: 2MASSW J0345 is a 
double-lined spectroscopic L dwarf system (Reid et al.\ 1999), and 
DENIS-P J1228 was shown to be double in HST imaging observations 
(Mart\'{\i}n et al.\ 1999). The latter is composed of equal-luminosity 
components with a projected separation 
of 0.275$''$ (5 AU at the 18 pc distance of DENIS-P J1228). 

Here we present the first results of a Keck near-infrared imaging survey of a
large sample of L dwarfs. At a projected distance of a few AU, 
our program is capable of detecting companions with luminosity 
similar to the primary. At further projected
distances, our survey is sensitive to
objects which are several magnitudes fainter than the methane brown dwarf,
Gl~229B.  In this work, we report the K-band detection of three L dwarf 
binaries, including DENIS-P J1228.

\section{Observations}

Our target sample was culled from the 2MASS and DENIS near-infrared sky 
surveys and consisted of objects spectroscopically confirmed to be L dwarfs.
We also included observations of a smaller sample of nearby very 
late M dwarfs. Imaging was carried 
out at the Keck I telescope with NIRC, a cryogenically-cooled near-infrared 
camera that incorporates a 256$\times$256 Indium-antimonide array at the 
f/25 focus in an optical framework which yields a 0.15$''$ plate scale and 
38$''$-square field of view (Matthews \& Soifer 1994). 
One-minute integrations were taken in the 
K-band filter at each of nine dithered positions separated by 5$''$. 
Individual frames 
were flat-fielded and differenced, then shifted and combined to create a 
sensitive composite image suitable for detecting companions to a limiting 
magnitude of m$_{\rm K} \approx 20$. At this level of sensitivity, several  
additional sources were typically detected in each frame. 
Repeat observations in a second epoch were taken one year or more later 
to determine if any of these share  
common proper motion with the target; second-epoch observations are complete 
for only a subset of the sample which includes 10 L dwarfs at present. 
Analysis of the completed survey will be presented in a future work.

In addition to the common proper motion analysis of faint sources, we
inspected the core of each of the primaries to search for extended emission 
associated with a marginally resolved binary. 
Second-epoch observations not only served to 
provide evidence of common proper motion, but also helped to ensure 
that any elongation was not due to time-dependent errors in 
phasing of the segmented primary mirror. Point-like sources 
observed nearby in the sky and within an hour of the target observations 
were chosen to serve as psf measurements. Dithered images of candidate 
binaries and psf stars were not shifted and combined but were treated as 
independent data sets. Psf stars were fit in duplicate to each of the 
candidate binary images using a least-squares minimization method.
Typically, nine psf frames were fit to each of nine image frames for a total 
of 81 independent fits. Properties of the
psf and target sources used in this work are listed 
in Table 1; results of the psf-fitting
are given in Table 2.

\section{Results}

Three objects met our 
critieria for reliable identification of a true close binary system,
i.e., the
availability of suitable psf observations and consistent results 
in psf-fitting for at least two epochs.
Contour plots of DENIS-P J1228, DENIS-P J0205, 
and 2MASSW J1146 are displayed in Fig.\ 1, together with the 
psf stars used to decompose them into separate components. The results of 
psf-fits in each epoch are listed in Table 2 and plotted in Fig. 2 and 3. 
Parameter estimates are consistent between two epochs; variations in the 
uncertainties are largely due to different seeing conditions. Conservatively, 
we state here the mean of the measurements in all epochs (rather than simply 
the best-seeing epoch) together with an uncertainty calculated as the root 
mean square difference from the mean. For the separation and PA of the 
component positions, these are plotted as large symbols in Fig.\ 2 and 
are $0.27\pm0.03''$, $0.51\pm0.03''$, $0.29\pm0.06''$ and $33\pm15^\circ$, 
$92\pm18^\circ$, $206\pm19^\circ$ 
for the component separations and PA's of DENIS-P J1228,
DENIS-P J0205, and 2MASSW J1146, respectively.  Projected separations 
correspond to physical separations of 4.9, 9.2, and 7.6 AU at distances 
implied by parallaxes listed in Table 1. Histograms of the flux-component 
ratios are plotted in Fig.\ 3; mean values are $1.1\pm0.4$,  
$1.0\pm0.4$, and $1.0\pm0.3$. 

\noindent {\it DENIS-P J1228} 

DENIS-P J1228 was discovered by Delfosse et al.\ (1997) and shown to have
lithium by Tinney et al.\ (1997) and Mart\'{\i}n et al.\ (1997). It is
an L5 V dwarf in the classification scheme of Kirkpatrick et al.\ (1999a)
with a temperature less than 1700 K implied by the absence of
Vanadium Oxide in its spectrum. Mart\'{\i}n et al.\ (1999) first discovered
the binary nature of DENIS-P J1228 and reported B/A flux ratios of
0.83, 0.86, and 0.87 in the HST/NICMOS filters, F110M, F135M, and F165M,
respectively. Our K-band flux ratio for the same components, $1.1\pm0.4$,
is consistent with a trend in the HST fluxes which implies 
that DENIS-P J1228B is slightly cooler, as is the value obtained on 
the best night of seeing, $0.99\pm0.25$.

The separation and PA reported by Mart\'{\i}n et al.\ (1999) are plotted
in Fig.\ 2 and were obtained in observations taken mid-way between 
Keck observations spaced one year apart. The results
clearly agree over an observational period
during which the system should have moved 0.21$''$ along PA 143.8$^\circ$
according to measurements by Dahn et al.\ (1999) that are listed
in Table 1. Indeed, we measure the mean proper motion with respect
to background stars in our field as $\mu$ = $0.23\pm0.05''$
at PA $130.4\pm8^\circ$. If either binary
component is a background star, this would result
in a later-epoch separation that is too large to be consistent with our 
observations.  Thus DENIS-P J1228 is a true common proper motion binary.

A recently determined value of the trigonometric parallax for DENIS-P J1228 
is listed in Table 1 as 0.0553$\pm$0.0029$''$ and implies a distance of 
$18.1\pm1.0$ pc. This result agrees with the distance estimate
of Mart\'{\i}n et al.\ (1999), based on
the spectroscopic parallax of Delfosse et al.\ (1997), and from  
which they derived a projected physical separation of 5 AU. Assuming
circular orbits and masses of 0.05 M$_\odot$, they calculate that orbital
motion can be detected by HST in one year. A precise determination
of dynamical mass will require observations over a much larger fraction of 
an orbit, however. The orbital period is at least 35 years, so a highly
refined mass estimate may not be forthcoming on timescales
competitive with analysis of closer binaries discovered in the interim.
Nevertheless, this binary has the smallest 
projected physical separation
of those presented here, hence, the best opportunity for near-term 
estimates of the dynamical mass. The importance of such an observation
is highlighted by the fact that, to date, secure dynamical mass
determinations have barely begun to extend to masses below the
hydrogen-burning limit (cf.\ Henry et al.\ 1999).

\noindent {\it DENIS-P J0205}

DENIS-P J0205 was first discovered by Delfosse et al.\ (1997) and
is reported in Kirkpatrick et al.\ (1999a) as a prototype of the L7 V
spectral class. Although no lithium line is evident in its spectrum,
this may indicate a sufficiently cool temperature to foster depletion
of atomic Li condensing out as LiCl. It is, in any case, the coolest
dwarf to date for which a similar or less luminous companion has been 
detected.

The proper motion measured for DENIS-P J0205, 
0.442$''$ at PA 82.6$^\circ$ (Table 1), is directed almost entirely
along the 92$^\circ$ PA derived for the binary and would
have produced a total motion of about 0.6$''$ in the $\sim1.5$ years
between observations reported here. This is greater than the binary separation
itself, which is measured consistently at both epochs to be 0.51$''$. 
Relative to 3 background stars
detected in our field of view, we measure the total proper motion of the 
system to be $0.58\pm0.22''$ along PA $77\pm7^\circ$ between epochs.
Consequently, it is securely established that DENIS-P J0205
comprises a common-proper-motion pair.
The trigonometric parallax, 0.0555$\pm$0.0023, implies
a distance of 18 pc, similar to that of DENIS-P J1228, and a projected
separation of 9.2 AU for the binary components.

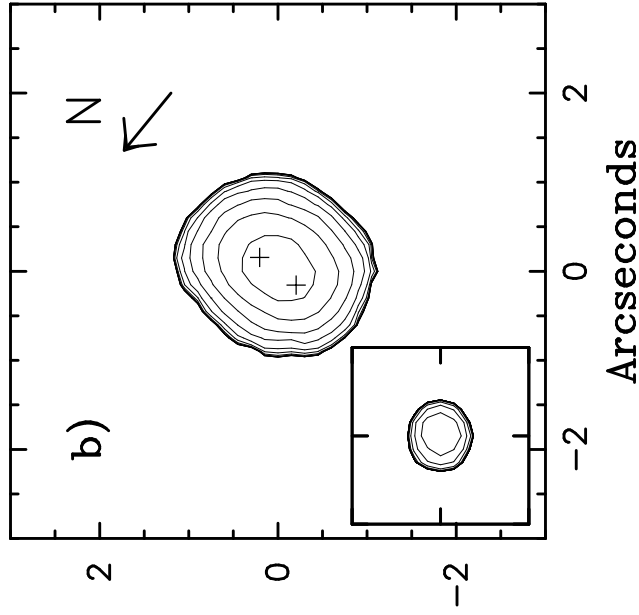
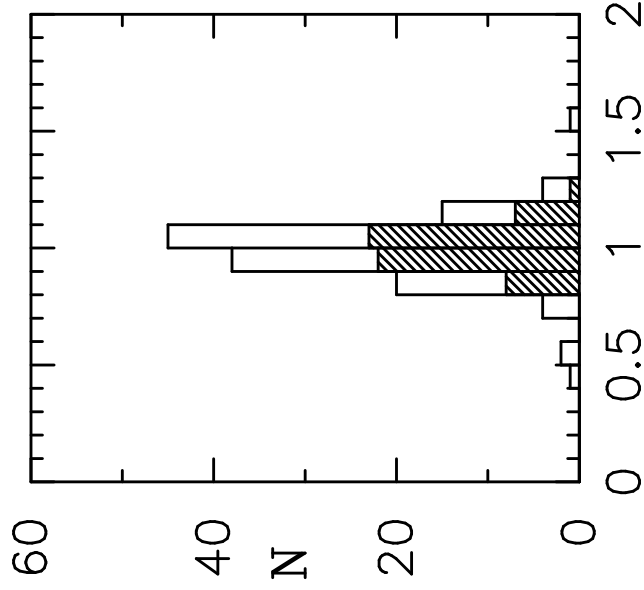
\noindent{\it 2MASSW J1146} 

The lithium dwarf, 2MASSW J1146, is classified L3 V. With a trigonometric
parallax $\pi$ = 0.0382$\pm$0.0013$''$, it is the most distant of the three
(D = 26.2 pc).  Its angular separation, 
$0.29\pm0.06''$ at PA $206\pm19^\circ$, implies a projected distance 
of 7.6 AU. It has the smallest proper motion of the three, 0.097$''$
at PA 20.2$^\circ$, but in a direction that is nearly anti-parallel 
to the PA of binary separation.  Consequently, differential proper motion 
would have been easily detected after the nearly 1-year interval separating
our observations. Second-epoch observations had relatively poor seeing;
the resulting scatter in Fig.\ 2 is evident. Nevertheless, the results 
still cluster about a mean coincident with that obtained at the
earlier epoch, indicating common proper motion.

For 2MASSW J1146, a second set of images was acquired in the first epoch 
using NIRC'S ``CH$_4$'' filter centered at  
$\lambda_{eff}$ = 2.269 $\mu$m with $\Delta\lambda$ = 0.155$\mu$m.
Binary separation and PA estimates from fits to these data are averaged
into the results reported above. Since the CH$_4$ flux component ratio 
agrees with the K-band ratios, within the uncertainties, we report here
the average of all three: 1.0$\pm$0.4. The absence of any difference 
precludes a contrasting presence 
of CH$_4$ absorption in one of the components,
as expected for a luminosity ratio of 1.

\section {Discussion}

The three binary systems presented here have similar projected separations
(5 to 9 AU) and luminosity ratios near unity. They represent the first
binary detections in preliminary analysis of a larger dual-epoch 
survey in which only 10 L-dwarf images have been completely analyzed 
in two epochs. No companions with wider separation or more highly
contrasting luminosities have been found thus far. Although analysis
of our complete sample is required to make unassailable statements about
the L dwarf binary distribution, these preliminary results suggest a
conjecture for further testing: namely, that multiple systems are not uncommon
in the L dwarf population, that their distribution peaks at radial separations
like that of both Jovian planets in our solar system and binary stars
generally ($\sim5-30$ AU), and that low-contrast mass ratios are 
common. The latter claim is especially in need of testing, since
our survey is not very sensitive to companions at the separations 
reported here if they have high luminosity contrast ratios. Further,
the magnitude-limited surveys from which our sample is taken are biased
toward the detection of equal-luminosity binaries,
since their combined luminosity is greater than for single stars of the same
spectral type.  Ultimately, techniques with both high resolution and high 
dynamic range must be applied to a volume-limited sample to reliably
identify the distribution of circumstellar bodies that encircle the
members of this population of very cool objects.

\acknowledgments

We thank I.N. Reid for reading an early draft of the
manuscript and for useful comments about the mass function of 
L dwarfs in the field.
mass function. For useful discussions, we also thank
E. Mart\'{\i}n with respect to background on DENIS-P J1228 and 
T.J. Henry regarding the practicalities of
dynamical mass determinations. 
We also wish to thank NIRC instrument specialists
Robert Goodrich and Randy Campbell, and operator assistants  
Ron Quick, Joel Aycock, Wayne Wack, and Chuck Sorenson. 
Data presented herein were obtained 
at WMKO, which is operated as a scientific partnership 
between Caltech, University of California, and NASA, and 
was made possible by the generous financial support of the
W.M. Keck Foundation.

\small
\hskip -0.5truein{
\begin{deluxetable}{lccccccc}
\small
\tablewidth{0pc}
\tablecaption{}
\tablehead{
\colhead{Name} & 
\colhead{$\alpha$ (J2000)} &
\colhead{$\delta$ (J2000)} & \colhead{ Sp T} &
\colhead{$K_s$} & \colhead{$\pi_{trig}('')$$^1$} &
\colhead{$\mu ('')$/yr$^1$} & \colhead{$\theta$(deg)$^1$}
}
\startdata
DENIS-P J0205 & $02^h 05^m 29.47^s$ & $-11^\circ$ 59$'$ 29.6$''$ &
L7 V & 12.99$\pm$0.04  &   0.0555$\pm$0.0023 &  0.442 &  82.6 \nl
LP 647-13 & $01^h 09^m 50.90^s$ & $-03^\circ$ 43$'$ 26.0$''$ &
M9 V & -- &  -- &  0.359 &  85 \nl
LHS 2351 & $10^h 16^m 35.00^s$ & $+27^\circ$ 51$'$ 48.0$''$ &
M8 V & 11.34$\pm$0.04  &  0.0461$\pm$0.0031 & 0.49 & 318.5 \nl
2MASSW J1145 & $11^h 45^m 57.20^s$ & $+23^\circ$ 17$'$ 30.0$''$ &
L1.5 V & 13.87$\pm$0.08  &  -- & -- & -- \nl
2MASSW J1146 & $11^h 46^m 34.50^s$ & $+22^\circ$ 30$'$ 53.0$''$ &
L3 V & 12.63$\pm$0.04  &   0.0382$\pm$0.0013 &  0.097 &  20.2 \nl
DENIS-P J1228 & $12^h 28^m 13.80^s$ & $-15^\circ$ 47$'$ 11.0$''$ &
L5 V & 12.81$\pm$0.03  & 0.0553$\pm$0.0029 &  0.21 &  143.8 \nl
KELU-1 & $13^h 05^m 40.20^s$ & $-25^\circ$ 41$'$ 06.0$''$ &
L2 V & 11.81$\pm$0.03  &   0.0516$\pm$0.0022 &  0.294 &  270.6 \nl
\tablerefs{(1) Dahn et al.\ (1999)}
\enddata
\end{deluxetable}
}

\hskip -0.5truein{
\begin{deluxetable}{lllcccc}
\small
\tablewidth{0pc}
\tablecaption{PSF Fitting Results}
\tablehead{
\colhead{Object} & \colhead{Date} &
\colhead{PSF} & 
\colhead{PSF} & \colhead{Sep} & \colhead{PA} &
\colhead{Flux}\\
\colhead{ } & \colhead{Observed} &
\colhead{STAR} & 
\colhead{FWHM} & \colhead{(arcsec)} & \colhead{(deg)} &
\colhead{Ratio}
}
\startdata
DENIS-P J0205 &  29Jul97 & LP 647-13 &
0.33$''$ & $0.51\pm0.02$  & 106$\pm5$ &  $0.99\pm0.08$ \nl
DENIS-P J0205 &  24Jan99 & LP 647-13 &
0.45$''$ & $0.51\pm0.04$  & 72$\pm10$ &  $1.00\pm0.26$ \nl
2MASSW J1146 &  14Feb98 & 2MASSW J1145 &
0.24$''$ & $0.30\pm0.01$  & 196$\pm3$ &  $0.87\pm0.07$  \nl
2MASSW J1146$^{a}$ &  16Feb98 & 2MASSW J1145 &
0.36$''$ & $0.28\pm0.03$  & 225$\pm10$ & $1.03\pm0.20$  \nl
2MASSW J1146 &  24Jan99 & LHS 2243 &
0.40$''$ & $0.29\pm0.10$  & 197$\pm24$ & $0.98\pm0.47$  \nl
DENIS-P J1228 & 14Feb98 & Kelu-1 & 
0.25$''$ & $0.25\pm0.02$  & 45$\pm6$ &  $1.28\pm0.50$  \nl
DENIS-P J1228 &  09Feb99 & Kelu-1 &
0.28$''$ & $0.30\pm0.02$  & 17$\pm7$ & $0.99\pm0.25$  \nl
\tablenotetext{a}{These observations were taken in NIRC's CH$_4$ filter
centered at $\lambda_{eff}$ = 2.269 $\mu$m with $\Delta\lambda$ = 0.155$\mu$m.}

\enddata
\end{deluxetable}\noindent

}

\figcaption{a) Contour plot of K-band imaging of DENIS-P J1228 together with
that of Kelu-1, the ``psf'' star used to derive binary component parameters. 
Contours are at logarithmic intervals. Crosses mark the separation and PA of
components derived in the psf fits to the data shown here. 
b) Plots as in a) for DENIS-P J0205 and
associated psf star, LP 647-13, c) Plots as in a) and b) for 2MASSW J1146
and psf star 2MASSW J1145.}

\figcaption{Binary separation and position angle from psf-fits 
of individual frames for 2MASSW J1146 (small open circles), DENIS-P J1228
(small open triangles), and DENIS-P J0205 (small filled triangles). The mean
of each of the measurements is plotted as a large open symbol for each
object with error bars that mark the rms deviation about the mean. The
HST/NICMOS result for DENIS-P J1228 is plotted as an open diamond.}

\figcaption{Histograms of the best-fit values of the flux component
ratio. Data for all epochs are plotted together as open bars. The
subset of measurements from 
the epoch with best seeing is plotted as hashed
bars. Component ratios are for B/A. For DENIS-P J1228, the identification
of the primary is taken from Mart\'{\i}n et al.\ (1999) at shorter 
wavelengths.}

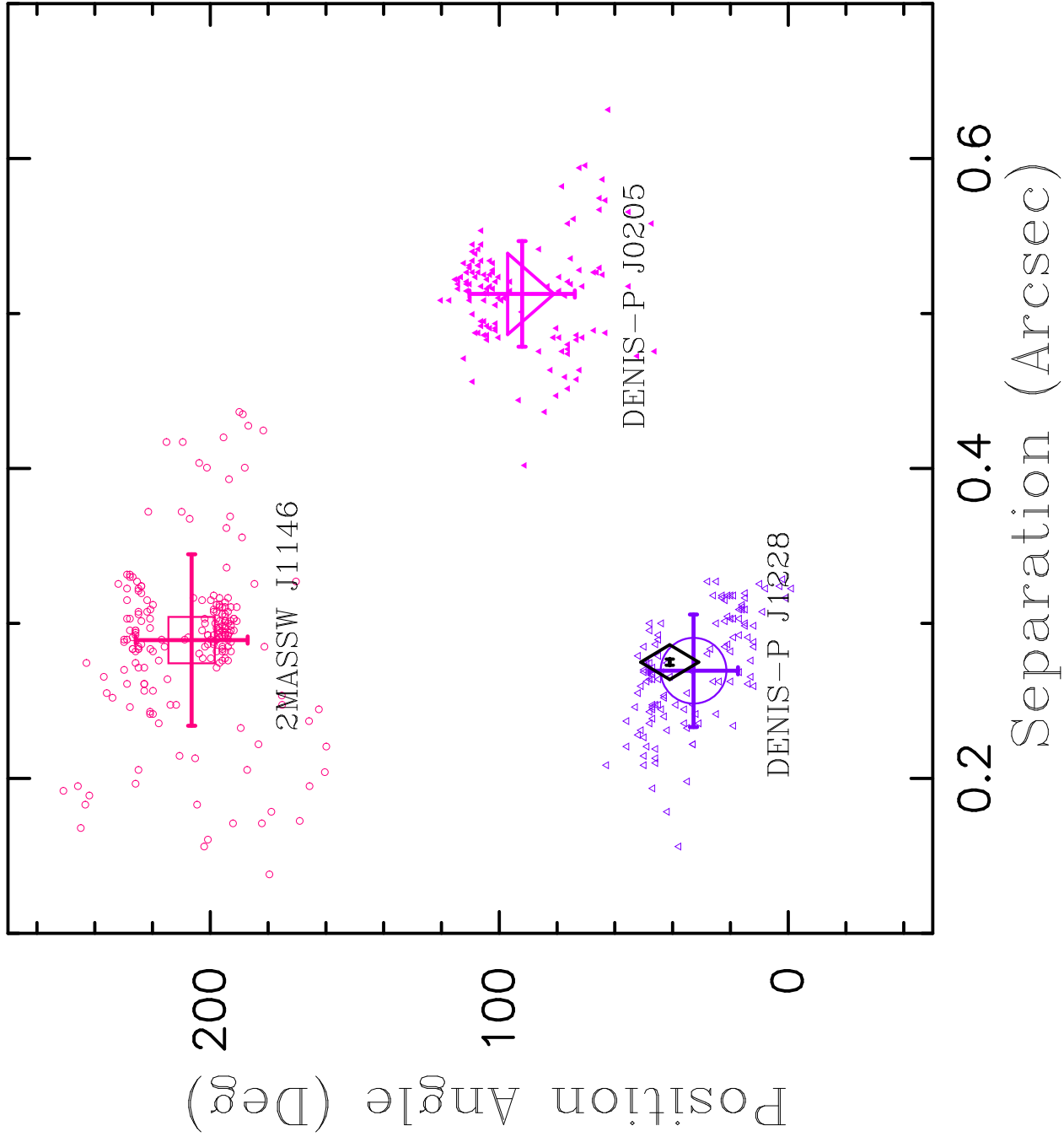
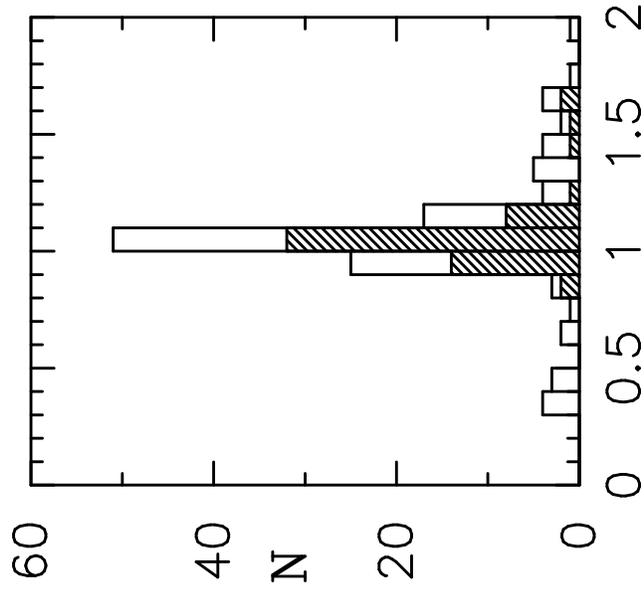
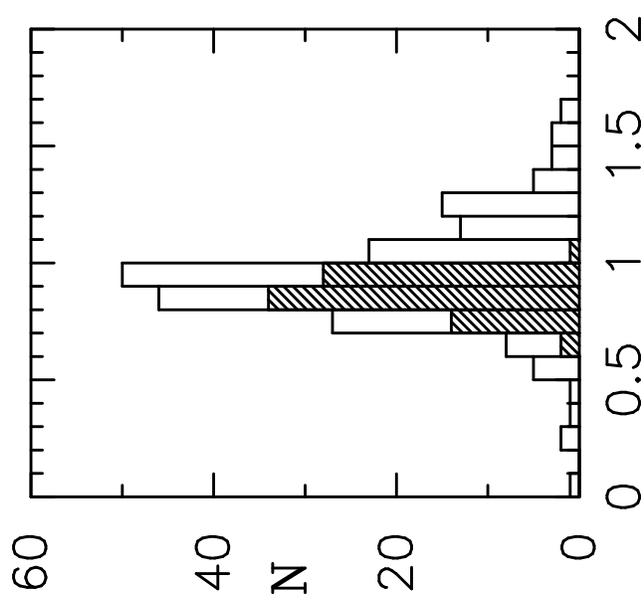
\end{document}